\documentstyle[12pt]{article}
\newtheorem{theorem}{Theorem}

\newtheorem{example}{Example}

\newcommand{\qed}{\hfill \raisebox{-1ex}{$\diamondsuit$}}
\newcommand{\QED}{\hfill \raisebox{-1ex}{$\diamondsuit$}}

\begin{document}
\baselinestretch
\newcommand{\proof}{\noindent {\bf Proof}\ \ }

\title{On binary constructions of quantum
 codes} 
\author{{\bf G\'erard Cohen}
\footnote{Ecole Nationale Sup\'erieure des T\'el\'ecommunications,
46 rue Barrault, 
75634 Paris, France;
e-mail: cohen@inf.enst.fr}, 
        {\bf Sylvia Encheva}\footnote{
Stord/Haugesund College
Skaareg. 103, 
5500 Haugesund,
Norway; 
e-mail: sbe@hsh.no}, 
        {\bf Simon Litsyn}
        \footnote{DIMACS, on sabbatical leave from Department of
                  Electrical Engineering-Systems, Tel Aviv University,
                  69978 Ramat Aviv, Israel; e-mail:
litsyn@eng.tau.ac.il
}}
\date{}

\maketitle

\begin{abstract} 
We improve estimates on the parameters of quantum codes obtained
by Steane's construction from binary codes. This yields several
new families
of quantum codes. 
\end{abstract}

\section{Introduction}

Although quaternary constructions \cite{cald98} yield
good quantum codes, building quantum codes from
binary ones results in efficient and easier
to implement families.
Such an approach was suggested by Calderbank and Shor \cite{cald96}
and Steane \cite{stea96a,stea96b}. Recently, Steane \cite{stea98}
proposed an enlargement of the Claderbank-Shor-Steane construction,
leading to several families of codes with fixed minimum distance and 
growing
length. In this paper, we further improve the estimates
of codes parameters obtained from Steane's construction,
present examples of new codes, and analyze
asymptotical non-constructive bounds. 

An (additive stabilizer) quantum code 
of length $n$
with $2^{k}$ codewords and minimum distance $d$, is an eigenspace of a 
commutative subgroup of the group $E$ of tensor products of Pauli
matrices. 
The commutativity condition is
\[ H_{x}.H_{z}^{T} + H_{z}.H_{x}^{T} = {\bf 0}, \]
where $H_{x}$ and $H_{z}$ are $(n-k) \times n$ binary matrices which 
together form the {\em stabilizer} ${\cal H} = (H_{x}|H_{z})$. 
The $2n$-vectors $(u_{x}|u_{z})$   
satisfying $H_{x}.u_{z} + H_{z}.u_{x} = 0$ form the code ${\cal C}$. 
These vectors are generated by ${\cal G} = (G_{x}|G_{z})$, where  
\[ H_{x}.G_{z}^{T} + H_{z}.G_{x}^{T} = {\bf 0}. \]
Define the {\it generalized weight}
of a vector $(u_{x}|u_{z})$ as the Hamming weight 
of the bitwise OR of $u_{x}$ and $u_{z}$. 
Let ${\cal C}^{\perp}$ be the dual of ${\cal C}$
with respect to the inner product 
$((u_{x}|u_{z}), (v_{x}|v_{z})) = u_{x}.v_{z} + u_{z}.v_{x}$.
The minimum (quantum) distance $d$ of the quantum code ${\cal C}$ is the 
largest 
generalized weight  
of a vector in ${\cal C} \setminus {\cal C}^{\perp}$. This code has
parameters $[[n,K,d]]$ where $K$ is $n-\log_2 |{\cal C}|$. 

Let $C[n,k,d]$ ($C[n,k]$ if $d$ is irrelevant) denote a binary
linear 
code of length $n$, dimension $k$ and minimum distance $d$. For a survey
of the theory of codes see e.g. \cite{macw77}.

The following construction was proposed by Steane \cite{stea98}. 

\begin{theorem} \label{th:st1} \cite{stea98}
Let $C[n,k,d]$, $C^{\perp} \subseteq C$,
be a classical binary linear error correcting code 
with generator matrix $G$. Let $C$ be
a subcode of a code $C'[n,k'>k+1,d']$ with generator matrix
$\left( \begin{array}{c}  G \\ G' \end{array} \right)$. Then
$${\cal G}=\left( \begin{array}{cc} G & 0 \\
                           0 & G \\
                           G'& PG'
\end{array} 
  \right)$$
where $P$ is a fix-point free map (for instance a permutation),
generates
a quantum code of parameters 
$[[n, k + k^{\prime} - n, 
\ge \min \left(d, {\left\lceil\frac{3d'}{2}\right\rceil}\right)]]$. 
\end{theorem} 

\proof
We sketch the proof for self-completeness. 
Consider any non-zero 
combination ${\bf u}=(u_x|u_z)$ of rows of ${\cal G}$. 
If no rows of the submatrix 
$(G'|PG')$ are involved in the generation of ${\bf u}$, then the
generalized
weight of ${\bf u}$ is $\geq d$. Otherwise, since
$\left( \begin{array}{c}  G \\ PG' \end{array} \right)$ also
generates $C'$, $u_x$ and $u_z$ are two
distinct non-zero codewords from $C'$, both of Hamming weight at least
$d'$ and at Hamming distance at least $d'$ apart. Therefore, their
bitwise OR
is at least 
$\lceil\frac{3d'}{2}\rceil$. \qed

As an illustration of the previous theorem, we present below a
record-breaking code. We also provide a table of a few
quantum codes which, although they just equal previously known codes
in terms of parameters, present the advantage of being obtained through
binary codes. The generator matrices of the corresponding $C'$'s 
can be found in the appendix.


\begin{example}
{\rm We extend the QR $[18, 9, 6]$-code $C$ to a 
$C'[18, 12, 4]$ with generator matrix written below. Applying Steane's
construction we obtain a $[[18, 3, 6]]$-code. The highest achievable
minimal 
distance $d$ of a $[[18, 3, d]]$-code is 5 or 6 \cite{cald98}. This
shows that $d=6$,thus closing the gap on $d$. \\

\[\begin{array}{l}

1 1 1 1 1 1 0 0 0 0 0 0 0 0 0 0 0 0\\
0 0 1 0 1 1 1 0 0 0 0 0 1 0 1 0 0 0\\
0 0 1 0 0 1 0 1 1 0 0 0 1 1 0 0 0 0\\
0 1 1 1 0 0 0 0 0 1 0 0 1 1 0 0 0 0\\
0 1 1 0 1 0 0 0 0 0 0 0 0 1 1 1 0 0\\
0 1 1 0 0 1 0 0 1 0 1 0 0 0 1 0 0 0\\
0 0 1 0 0 0 0 0 1 1 0 0 1 0 1 0 1 0\\
0 0 1 0 0 0 0 0 1 1 0 0 0 1 0 1 0 1\\
0 1 1 0 0 0 0 0 0 1 1 1 0 0 0 1 0 0\\
1 1 0 0 0 0 1 0 1 0 0 0 0 0 0 0 0 0\\
0 0 0 0 0 0 0 0 1 0 0 0 1 1 0 0 0 1\\
0 0 0 0 0 0 0 1 0 1 0 0 0 0 1 0 0 1\\
\end{array} \]

}
\end{example}

\begin{table}
\begin{center}
\caption{Some quantum codes obtained by theorems $^{~}$\ref{th:st1},
$^{~}$\ref{th:st2}}

\begin{tabular}{|r|c|l|l|l|l|l|c|} \hline
$n$ &$k$&$k^{\prime}$&$d$&$d^{\prime}$&$K$&$d_{[[n,k]]}$& Remarks  \\                                                                                      \hline
 8  & 4 &   7        & 4 &   2        &3  & 3            & \\
12  & 6 &   10       & 4 &   2        & 4 & 3            &       \\
12  & 6 &   11       & 4 &   2        & 5 & 3            &       \\
14  & 7 &   9        & 4 &   2        & 2 & 4            &
$d_2^{\prime}>\lceil\frac{3d'}{2}\rceil$, apply Th $^{~}$\ref{th:st2}\\
14  & 7 &   10       & 4 &   2        & 3 & 4            & 
$d_2^{\prime}>\lceil\frac{3d'}{2}\rceil$, apply Th $^{~}$\ref{th:st2}\\
18  & 9 &   12       & 6 &   4        & 3 & 6            & optimal\\

\hline
\end{tabular}
\label{t41b}
\end{center}
\end{table}

In Table 1, $K$ and $d_{[[n,k]]}$ are the dimension and the minimum
distance of a quantum code
obtained by theorems $^{~}$\ref{th:st1}, $^{~}$\ref{th:st2} .\\

\section{An improvement}

We need the following notion introduced by Wei \cite{wei91}.
The $i$-th {\it generalized distance} $d_i$ of a linear code $C[n,k]$ 
is the minimum size 
of the support of a i-th dimensional subcode of $C$.
For the best known bounds on generalized weights consult 
\cite{cohe94,hell95,tsfa95,wei91}.

Clearly, for $i=1$, $d_1$ is the minimum distance of the code;
for $i=2$, $d_2$ is the minimum weight of the bitwise OR of two
{\it different} nonzero codewords. It follows easily from the Griesmer
bound 
(see e.g. \cite[Chap.17]{macw77}), 
that $d_2 \geq \lceil 3d/2 \rceil $. We are ready now for the
improvement.

\begin{theorem}
\label{th:st2}
With the notation of the previous theorem, $\cal G$ generates a 
quantum code of parameters $[[n, k+k'-n, min(d, d'_2)]]$, where
$d_2'$ is the second generalized distance of $C'$.
\end{theorem}

\proof
Recalling the proof of 
the previous theorem, the only thing left to be shown
is that if rows of $(G'|PG')$ are involved in the generation of ${\bf
u}$, then
the generalized weight of ${\bf u}$ is 
$\geq d'_2$. In this case both $u_x$ and $u_z$ are in
$C'$, and distinct since $P$ has no fix point,  and the assertion
follows
from the definition of the generalized distance.
This bound is at least as strong as the previous one, since 
$d'_2 \geq 3d'/2$ 
by the
remark preceding the theorem.
\qed

\section{Some new quantum codes}

Steane \cite{stea98} proved that the primitive BCH codes of length
$2^m-1$
contain their 
duals if and only if their designed distance $d=2t+1$ satisfies $d \le
2^{
\lceil m/2 \rceil}-1$. It follows from \cite[Corollary 8, Chapter
9]{macw77}
that in this case the codes have parameters
$[2^m-1,2^m-1-mt,2t+1]$. Moreover, these codes are nested, i.e. form
a chain for the inclusion relation when $t$ increases.
Extending them with a parity bit, we 
derive the following families using Theorem \ref{th:st1}:

\begin{eqnarray*}
F_0\ (d=6 \ell, d'=4 \ell) & [[2^m,2^m-(5\ell-2)m-2,6\ell]] & \mbox{for\
}
6 \ell \le 2^{\lceil m/2 \rceil} ;\\
F_2\ (d=6 \ell+2, d'=4 \ell+2) &  [[2^m,2^m-5\ell m -2,6\ell+2]] &
\mbox{for\ }
6 \ell+2 \le 2^{\lceil m/2 \rceil} ;\\
F_3\ (d=6 \ell+4, d'=4 \ell+2) & [[2^m,2^m-(5\ell+1)m-2,6\ell+3]] & 
\mbox{for\ }
6 \ell +4 \le 2^{\lceil m/2 \rceil} .
\end{eqnarray*}
 
Theorem 6 of \cite{cald98} shows how to construct an $[[n,K+1,d-1]]$
code
from an $[[n,K,d]]$ code. Using it we construct from $F_0$ the following
family:
$$F_5 \quad  [[2^m,2^m-(5\ell+3)m-1,6\ell+5]] \quad \mbox{for\ }
6 \ell+6 \le 2^{\lceil m/2 \rceil}.$$

It is tempting to conjecture the existence of families of codes with
parameters
$$F_a=[[2^m,2^m-(5 \ell+a-2)m+b,6 \ell+a]],$$
where $a=0,1,2,3,4,5$ and $b$ is a small integer constant.
Using Theorem \ref{th:st2} we construct such a family for $a=4$. The
case
$a=1$ remains open.

\begin{theorem}
For $6 \ell+4 \le 2^{\lceil m/2 \rceil}$ there exist quantum codes
with parameters
$$F_4=[[2^m,2^m-(5\ell+2)m-1,6\ell+4]] .$$
\end{theorem}

\proof Consider the extended BCH code,
$C[2^m,2^m-(3\ell+1)m-1,6\ell+4]$.
As the code $C'$ we take the union of the code $C_1=
[2^m,2^m-(2\ell+1)m-1,4\ell+4]$ with any coset of $C_1$, say $C_2$, in
the code $[2^m,2^m-2\ell m-1,4\ell+2]$. This union is a
$C'=[2^m,2^m-2(\ell+1) m,4\ell+2]$ code. 
However, $d'_2 \ge 6 \ell+4 > 3d'/2=6 \ell+3$.
Indeed, consider the bitwise OR of any two codewords of $C'$. If they
both have weight $4 \ell+2$ they belong to $C_2$, hence are at
least
at distance $4 \ell+4$ apart, yielding a generalized weight 
at least $6 \ell+4$. Otherwise, one of the words has weight at
least $4 \ell+4$, and even if the second one is of weight $4 \ell+2$,
since 
their distance is at least $4 \ell+2$, this again
guarantees 
a minimum generalized weight of at least $6 \ell+4$.
\qed
 
\section{Asymptotical behaviour}

Let us consider now the asymptotical-in $n$-
non constructive behavior of codes
obtained by use of Theorem \ref{th:st2}. 
Let $R_Q({\cal C})=K/n$ and $\delta_Q({\cal C})=d/n$ 
stand for the rate and relative minimum distance of ${\cal C}=
[[n,K,d]]$. We are interested in
$$R_Q(\delta_Q)= \limsup_{n \rightarrow \infty} R_Q({\cal C}),$$
where the limit is taken over all codes with $\delta_Q({\cal C})
\ge \delta_Q$.

The best known lower bound on $R_Q$ is obtained \cite{cald98}
via codes over $GF(4)$:
\begin{equation}
\label{vg4}
R_Q \ge 1-\delta_Q \log_2 3-H(\delta_Q) ,
\end{equation}
where
$H(x)= -x log_2 x - (1-x) log_2 (1-x)$ is the binary entropy function.
For the best upper bounds see \cite{ashi99}.

For binary constructions, Calderbank and Shor \cite{cald96} proved
a weaker bound

\begin{equation}
\label{vg2}
R_Q^b \ge 1-2 H(\delta_Q).
\end{equation} 

It is fairly easy to show that Theorem \ref{th:st1} yields
\begin{equation}
\label{sta}
R_Q^b \ge 1-H(\delta_Q)-H(2 \delta_Q/3) ,
\end{equation} 
which is better than (\ref{vg2}).

\begin{theorem}
$$R_Q^b \ge 1-\frac{\delta_Q \log_2 3}{2} -3\frac{H(\delta_Q)}{2}.$$
\end{theorem}

\proof It follows along the lines of the standard proofs,  
see e.g. \cite[Section 5]{cald96}.  
Let $k=\lfloor Rn \rfloor, k'=\lfloor R'n \rfloor, 
d = d'_2=\lfloor \delta_Q n \rfloor$ 
be the parameters of $C$ 
and $C'$. For given $n$, $k$ and $k'$, consider two families of codes:
$${\cal C}^{(1)}= \{C[n,k] \mbox{\rm \ such that\ } 
C^{\perp} \subseteq C\},$$
$${\cal C}^{(2)}= \{C'[n,k']  \mbox{\rm \ such that\ } C \subset C'
\mbox{\rm \ for at least one \ } C \in {\cal C}^{(1)}\}.$$

The proof consists of three steps.
We first prove that almost all codes in ${\cal C}^{(1)}$ lie above the
Varshamov
Gilbert (VG) bound (see e.g. \cite{macw77}); 
then that almost all codes in ${\cal C}^{(2)}$ lie above an
analog of the VG bound for the second generalized distance; finally we
combine
these two results to construct binary quantum codes satisfying the
theorem. 
Arguments
from \cite{cald96} yield that 

\noindent a) every non-zero vector of even weight belongs to the same
number,
say A, of
codes from ${\cal C}^{(1)}$;

\noindent b) every pair of non-zero vectors of even weight belongs to
the
same number, say B, of codes from ${\cal C}^{(2)}$;

\noindent c) every code in ${\cal C}^{(1)}$ belongs to the same number
of codes in ${\cal C}^{(2)}$.

By hypothesis a),

$$(2^{n-1}-1)A=(2^k-1) |{\cal C}^{(1)}|,$$
and if
$$A \sum_{j=1}^{(d-1)/2} {n \choose 2j} \le A\frac{(2^{n-1}-1)}{n
(2^k-1)}=
|{\cal C}^{(1)}|/n$$
then at least $(1-1/n) |{\cal C}^{(1)}|$ codes from 
${\cal C}^{(1)}$ satisfy
the VG bound
\begin{equation}
\label{vg}
 R \ge 1-H(\delta_Q) . 
\end{equation}

Consider now the pairs of non-zero even-weight vectors with bitwise OR
equal
exactly to some vector of weight $t$. Their number is less than  
$$ \frac{1}{2} \sum_{j=0}^{[t/2]} {t \choose 2j} 2^{2j-1} 
=\frac{1}{8} ((1+2)^t+(1-2)^t) \le \frac{1}{8} (3^t+1) .$$
The total number of such pairs for $t \le d$ is at most 
$$\frac{1}{8} \sum_{t=1}^d {n \choose t} (3^t+1).$$
By argument b)

$${2^{n-1}-1 \choose 2} B={2^{k'}-1 \choose 2} |{\cal C}^{(2)}|,$$ 
and analogously to the previous argument we have that at least 
$(1-1/n) |{\cal C}^{(2)}|$ codes from ${\cal C}^{(2)}$ satisfy
\begin{equation}
\label{vg22}
R' \ge 1-\frac{\delta_Q \log_2 3}{2} -\frac{ H(\delta_Q)}{2}.
\end{equation}
 
Finally, argument c) implies the existence of a code satisfying 
both (\ref{vg}) and (\ref{vg22}).
By Theorem 2, this yields a binary quantum code $[[n,(R+R'-1)n,\delta_Q
n]]$
as stated in Theorem 4.
\QED

\pagebreak
\section{Appendix}

 Generator matrices for codes $C'$ in Table 1.\\

$C'[12,10,2]$ (two possibilities)\\

\[ \left(\begin{array}{c}
1 1 1 1 0 0 0 0 0 0 0 0\\
0 0 1 1 1 1 0 0 0 0 0 0\\
0 0 0 0 1 1 1 1 0 0 0 0\\
0 0 0 0 0 0 1 1 1 1 0 0\\
0 0 0 0 0 0 0 0 1 1 1 1\\
0 1 0 1 0 1 0 1 0 1 0 1\\
1 1 0 0 0 0 0 0 0 0 0 0\\
1 0 1 0 0 0 0 0 0 0 0 0\\
1 0 0 0 0 0 0 0 1 0 0 0\\
1 0 0 0 0 0 0 0 0 0 0 1
\end{array}\right), \]


\[ \left(\begin{array}{c}
1 1 1 1 0 0 0 0 0 0 0 0\\
0 0 1 1 1 1 0 0 0 0 0 0\\
0 0 0 0 1 1 1 1 0 0 0 0\\
0 0 0 0 0 0 1 1 1 1 0 0\\
0 0 0 0 0 0 0 0 1 1 1 1\\
0 1 0 1 0 1 0 1 0 1 0 1\\
0 0 0 0 0 0 0 0 1 0 1 0\\
0 0 1 0 0 0 0 1 0 0 1 0\\
0 0 1 0 1 0 0 0 0 1 0 0\\
1 1 0 0 0 0 0 0 0 0 0 0
\end{array}\right), \]


\newpage
$C'[14,9,2]$ \\

\[ \left(\begin{array}{c}

1 1 1 1 0 0 0 0 0 0 0 0 0 0\\
0 0 1 1 1 1 0 0 0 0 0 0 0 0\\
1 0 1 0 1 0 1 0 0 0 0 0 0 0\\
0 0 0 0 0 0 0 1 1 1 1 0 0 0\\
0 0 0 0 0 0 0 0 0 1 1 1 1 0\\
0 0 0 0 0 0 0 1 0 1 0 1 0 1\\
1 1 1 1 1 1 1 1 1 1 1 1 1 1\\
1 0 0 0 1 0 0 0 0 0 0 0 0 1\\
0 1 0 0 0 0 0 0 1 0 0 0 0 1\\
\end{array}\right), \]




$C'[14,10,2]$ \\

\[ \left(\begin{array}{c}
1 1 1 1 0 0 0 0 0 0 0 0 0 0\\
0 0 1 1 1 1 0 0 0 0 0 0 0 0\\
1 0 1 0 1 0 1 0 0 0 0 0 0 0\\
0 0 0 0 0 0 0 1 1 1 1 0 0 0\\
0 0 0 0 0 0 0 0 0 1 1 1 1 0\\
0 0 0 0 0 0 0 1 0 1 0 1 0 1\\
1 1 1 1 1 1 1 1 1 1 1 1 1 1\\
1 0 0 0 1 0 0 0 0 0 0 0 0 1\\
0 1 0 0 0 0 0 0 1 0 0 0 0 1\\
1 0 1 0 0 0 0 0 0 0 0 0 0 0
\end{array}\right), \]


\end{document}